\documentclass{article}[11pt]
\usepackage{amsfonts}
\usepackage{CJK,amsmath,indentfirst}
\usepackage{graphics}
\usepackage{graphicx}
\begin{document}

\title{\bf Nonlocal conservation laws and related B\"acklund transformations via reciprocal transformations}

\author{\footnotesize S. Y. Lou$^{1,2}$\thanks{Email: lousenyue@nbu.edu.cn}\\
\footnotesize $^{1}$ \it Shanghai Key Laboratory of Trustworthy Computing, East China Normal University, Shanghai 200062, China\\
\footnotesize $^{2}$\it Faculty of Science, Ningbo University, Ningbo, 315211, China
}
\date{}
\maketitle
\parindent=0pt
\textbf{Abstract:} A set of infinitely many nonlocal conservation laws are revealed for (1+1)-dimensional evolution equations. For some special known integrable systems, say, the KdV and Dym equations, it is found that different nonlocal conservation laws can lead to same new integrable systems via reciprocal transformations. On the other hand, it can be considered as one solution of the new model obtained via reciprocal transformation(s) can be changed to different solutions of the original model. The fact indicates also that two or more different (local and nonlocal) conservation laws can be used to find implicit auto-B\"acklund transformations via reciprocal transformation to other systems. 
\\

\vskip.4in
\renewcommand{\thesection}{\arabic{section}}
\parindent=20pt

Conservation laws play an important role to understand and even solve nonlinear systems \cite{Olver}. In this short letter, we try to give a special set of infinitely many nonlocal conservation laws for (1+1)-dimension evolution equations
\begin{equation}\label{EE}
v_t=K(v),
\end{equation}
with $K$ being an any given function of $\{x,\ t,\ v\}$ and the derivatives of $v$ with respect to $x$ by introducing some arbitrary freedoms to Lax pairs.

For simplicity, we only consider the following second order formal Lax pairs
\begin{subequations}\label{Lax}
\begin{equation}\label{Lx}
S\psi=0,\qquad S\equiv \partial_{x}^2-F(v),
\end{equation}
\begin{equation}\label{Lt}
Y\psi=0, \qquad Y\equiv \partial_t+q_x+2q\partial_x,
\end{equation}
\end{subequations}
where $F(v)\equiv F(x,\ t,\ v,\ v_x,\ v_{xx},\ \ldots) \equiv F$ is an arbitrary function of the space-time $\{x,\ t\}$, the field $v$ and its derivatives with respect to $x$.

We require the potential $q$ is related to the field $v$ by
\begin{equation}\label{pot}
Lq=F'K(v),\qquad L\equiv \partial_{x}^3-4F\partial_x-2(F'v_x),
\end{equation}
where $F'$ is the standard linearized operator, Gateaux derivative, defined by
\begin{equation*}
F'f\equiv \left.\frac{\partial}{\partial \epsilon}F(v+\epsilon f)\right|_{\epsilon=0}
\end{equation*}
for arbitrary $f$.

Now, it is straightforward to find that the
consistent condition of the Lax pair is just
\begin{equation}\label{SY}
\left.[S,Y]\psi\right|_{S\psi=Y\psi=0}=\psi F'(v_t-K(v)),\ [S,\ Y]\equiv SY-YS.
\end{equation}

In other words, if $q$ in \eqref{Lax} is defined by \eqref{pot}, then \eqref{Lax} with arbitrary $F$ is a (weak) Lax pair of the evolution equation \eqref{EE}.

It is interesting that \eqref{Lt} can be rewritten as
\begin{equation}\label{CL}
\rho_t+J_x=0,\quad \rho=\psi^{-2},\ J=2q\psi^{-2},
\end{equation}
which is just a conservation law.

The conservation law is nonlocal for the evolution equation \eqref{EE} because the conserved density $\rho$ is non-locally related to the field $v$ by three partial differential equations \eqref{Lax} and \eqref{pot}. However, it is a local conservation law for the equation system $\{\eqref{EE},\ \eqref{Lax},\ \eqref{pot}\}$.

On the other hand, \eqref{CL} is only one conservation law for the equation system $\{\eqref{EE},\ \eqref{Lax},\ \eqref{pot}\}$. However, \eqref{CL} implies infinitely many nonlocal conservation laws for the single evolution equation \eqref{EE} because $F$ is arbitrary.

It should be mentioned that many special Lax pairs are the particular cases of \eqref{Lax} for some well-known  integrable systems such as the KdV equation \cite{KdV},
modified KdV equation \cite{mKdV}, Dym equation \cite{Dym}, Camassa-Holm equation \cite{CH} and Ito system \cite{Ito}.

An important question is what can be obtained from the nonlocal conservation law \eqref{CL}? To answer this question, we restrict ourselves to some special cases because we have not yet good ideas for general evolution equation \eqref{EE}.

The first example is naturally the KdV equation
\begin{equation}\label{KdV}
v_t+v_{xxx}+6vv_x=0.
\end{equation}
For the KdV equation \eqref{KdV}, if we take $F=\lambda-v$, then the potential $q$ equation \eqref{pot} has a special solution $q=v+2\lambda$.
Thus, the conservation law \eqref{CL} becomes
\begin{equation}\label{CLkdv}
(\psi^{-2})_t=[-2(v+2\lambda)\psi^{-2}]_x.
\end{equation}
It is known that from every conservation law, we can construct a reciprocal transformation.

From \eqref{CLkdv} we can define a potential $z$
\begin{equation}\label{zxt}
z_x=\rho,\quad z_t=-2(v+2\lambda)\rho,\quad \rho=\psi^{-2}.
\end{equation}
Taking $\{z,\ t\}$ as new independent variables, i.e.,
\begin{equation}\label{pv}
v=v(z,\ t),\ \psi=\psi(z,\ t),
\end{equation}
we can obtain the well known Dym equation
\begin{equation}\label{Dym}
\rho_t+\rho^3\rho_{zzz}=0,
\end{equation}
while the field $v$ is given by
\begin{equation}\label{Dymv}
v=\frac34\frac{\rho_z^2}{\rho^4}
-\frac12\frac{\rho_{zz}}{\rho^3}+\lambda.
\end{equation}

On the other hand, for the KdV equation, related to the square eigenfunction symmetry, there is another nonlocal conservation law in the form
\begin{equation}\label{CLKdV}
\rho_{1t}=J_{1x},\quad \rho_1=\psi^{2},\ J_1=2(u-4\lambda)\rho_1+2\psi_x^2.
\end{equation}
In the same way, according to the conservation law \eqref{CLKdV}, we can define the potential $y$ in the form
\begin{equation}\label{ry}
y_x=\rho_{1},\ y_t=J_{1}.
\end{equation}
Taking $\{y,\ t\}$ as independent variables, i.e.,
\begin{equation}\label{pvy}
v=v(y,\ t),\ \psi=\psi(y,\ t),
\end{equation}
we find again the Dym equation
\begin{equation}\label{Dym1}
\rho_{1t}+\rho_1^3\rho_{1yyy}=0,
\end{equation}
while the field $v$ is determined by
\begin{equation}\label{Dymv1}
v=-\frac14{\rho_{1y}^2}
-\frac12\rho_1\rho_{1yy}+\lambda.
\end{equation}

On one hand, two different nonlocal conservation laws of the KdV equation lead to one same Dym equation. On the other hand one solution of the Dym equation leads to two different solutions of the KdV equation via two conservation laws. The fact implies that two different nonlocal conservation laws will lead to a B\"acklund transformation of the KdV equation.

To see whether this property holds for other models, we study the Dym equation,
\begin{equation}\label{HD}
v_{t}=v^3v_{xxx},
\end{equation}
which is only a re-notation of \eqref{Dym} for simplicity.

For the Dym equation \eqref{HD}, if we take
$$F=\frac{\lambda}{2v^2},$$
then the potential $q$ equation has a special solution
\begin{equation}\label{HDq}
q=\lambda v.
\end{equation}
Correspondingly, the conservation law \eqref{CL} becomes \begin{equation}\label{CLHD}
(\psi^{-2})_t=(2\lambda v \psi^{-2})_x.
\end{equation}
Thus we can introduce new independent variable $z$ via \begin{equation}\label{zxtHD}
z_x=\mu, \ z_t=2\lambda v\mu,\ \mu= \psi^{-2}.
\end{equation}
Under the new independent variables, we have
\begin{equation}\label{pvHD}
\psi=\psi(z,t),\ v=v(z,t).
\end{equation}
Substituting \eqref{pvHD} with \eqref{zxtHD} into the Dym equation \eqref{HD} yields a new equation
\begin{equation}\label{HDL}
\mu_t=\frac{(2\lambda)^{3/2}\mu^3\mu_{zzz}}
{(\mu_z^2-2\mu\mu_{zz})^{3/2}}
\end{equation}
while the solution of the Dym equation \eqref{HD} is related to $\mu$ by
\begin{equation}\label{HDLv}
v=\frac{\sqrt{2\lambda}}
{\sqrt{\mu_z^2-2\mu\mu_{zz}}}.
\end{equation}
It is not difficult to verify that the Dym equation possesses another nonlocal conservation Law
\begin{equation}\label{CLHD1}
(\psi^2v^{-2})_t=\big[2v^{-1}(2vv_x\psi\psi_x-2v^2\psi_x^2-v\psi^2v_{xx}+2\lambda \psi^2)\big]_x.
\end{equation}
Now, we introduce $y$ by
\begin{equation}\label{yxt1}
y_x=\psi^2v^{-2},\quad y_t=2v^{-1}(2vv_x\psi\psi_x-2v^2\psi_x^2-v\psi^2v_{xx}+2\lambda \psi^2)
\end{equation}
and suppose
\begin{equation}\label{pvHD1}
\psi=\psi(y,t),\ v=v(y,t).
\end{equation}
Substituting \eqref{pvHD1} with \eqref{yxt1} into the Dym equation \eqref{HD} leads the same equation \eqref{HDL} but with different variables
\begin{equation}\label{HDLt}
\theta_t=\frac{(2\lambda)^{3/2}\theta^3\theta_{yyy}}
{(\theta_y^2-2\theta\theta_{yy})^{3/2}},\
\end{equation}
while the solution of the Dym equation \eqref{HD} and the spectral function are related to $\theta$ by
\begin{subequations}
\begin{equation}\label{HDLv1}
v=\frac{\sqrt{2\lambda}\lambda}{\theta_y^2}
{\sqrt{\theta_y^2-2\theta\theta_{yy}}}.
\end{equation}
\begin{equation}\label{HDLp}
\psi=\lambda\sqrt{\theta}\theta_y^{-2}.
\end{equation}
\end{subequations}
Thus, as in the KdV case, the same conclusion is obtained for the Dym system: Two nonlocal conservation laws lead to a same new equation \eqref{HDL}. One solution of the new equation \eqref{HDL} leads to two solutions of the Dym equation \eqref{HD} via \eqref{HDLv} and \eqref{HDLv1}. The existence of two solutions of the Dym equation related to a same equation \eqref{HDL} implies a B\"acklund transformation for the Dym equation owing to the existence of two different nonlocal conservation laws. Similar procedure can be continued for the new equation \eqref{HDL} by investigating its nonlocal conservation laws.

In summary, by formally introducing a potential $q$ via \eqref{pot}, a weak Lax pair with an arbitrary function $F(v)$ is obtained for any (1+1)-dimensional evolution equations \eqref{EE}. Starting from the Lax pair, infinitely many nonlocal conservation laws are obtained. Various Lax pairs for known integrable systems are special cases of \eqref{Lax} with some special fixed $F(v)$. For these kinds of well known integrable systems, the new found conservation laws can be used to find new integrable systems via reciprocal transformations and B\"acklund transformations of the original model via the combination of other reciprocal transformations related to different nonlocal conservation laws. For general evolution equation, the nonlocal conservation laws may be not enough to guarantee the integrability. However, it must be useful to help to understand the special solutions of the nonlinear systems. It is worth to pay more attentions on this type of conservation laws.

The author is grateful to thank Professors C W Cao, D X Kong, Y Q Li, Y Chen, Q P Liu, X B Hu, R X Yao, E G Fan, C Z Qu, J S He and D J Zhang for their helpful suggestions and fruitful discussions. The work was sponsored by the National Natural Science Foundations of China (Nos. 11175092, 11275123, 11205092 and 10905038), Shanghai Knowledge Service Platform for Trustworthy Internet of Things (No. ZF1213) and K. C. Wong Magna Fund in Ningbo University.

\small{
}
\end{document}